\documentclass{pr-imfp04}

\def\lsim{\mathrel{\mathop
  {\hbox{\lower0.5ex\hbox{$\sim$}\kern-0.8em\lower-0.7ex\hbox{$<$}}}}}
\def\gsim{\mathrel{\mathop
  {\hbox{\lower0.5ex\hbox{$\sim$}\kern-0.8em\lower-0.7ex\hbox{$>$}}}}}

\begin{document}

\newcommand{\half}{{1\over2}}

\title{Modern Cosmology}

\author{Juan Garc\'{\i}a-Bellido}

\address{Departamento de F\'{\i}sica Te\'orica, Universidad Aut\'onoma 
de Madrid,\\ Cantoblanco, 28049 Madrid, Spain. \
E-mail: juan.garciabellido@uam.es}

\maketitle

\abstracts{In these notes I will review our present understanding of
the origin and evolution of the universe, making emphasis on the most
recent observations of the acceleration of the universe, the precise
measurements of the microwave background anisotropies, and the
formation of structure like galaxies and clusters of galaxies
from tiny primodial fluctuations generated during inflation.}

\section{Introduction}

The last five years have seen the coming of age of Modern Cosmology, 
a mature branch of science based on the Big Bang theory and the
Inflationary Paradigm. In particular, we can now define
rather precisely a Standard Model of Cosmology, where the basic
parameters are determined within small uncertainties, of just a few
percent, thanks to a host of experiments and observations.  This
precision era of cosmology has become possible thanks to important
experimental developments in all fronts, from measurements of
supernovae at high redshifts to the microwave background anisotropies,
and to the distribution of matter in galaxies and clusters of
galaxies.

In these lecture notes I will first introduce the basic concepts and
equations associated with hot Big Bang cosmology, defining the main
cosmological parameters and their corresponding relationships.  Then I
will address in detail the three fundamental observations that have
shaped our present knowledge: the recent acceleration of the universe,
the distribution of matter on large scales and the anisotropies in the
microwave background. Together these observations allow the precise
determination of a handful of cosmological parameters, in the context
of the inflationary plus cold dark matter paradigm.

\section{Big Bang Cosmology}

Our present understanding of the universe is based upon the successful
hot Big Bang theory, which explains its evolution from the first
fraction of a second to our present age, around 13.6 billion years
later. This theory rests upon four robust pillars, a theoretical
framework based on general relativity, as put forward by Albert
Einstein\cite{Einstein} and Alexander A. Friedmann\cite{Friedmann} in
the 1920s, and three basic observational facts: First, the expansion of
the universe, discovered by Edwin P. Hubble\cite{Hubble} in the 1930s,
as a recession of galaxies at a speed proportional to their distance
from us. Second, the relative abundance of light elements, explained by
George Gamow\cite{Gamow} in the 1940s, mainly that of helium, deuterium
and lithium, which were cooked from the nuclear reactions that took
place at around a second to a few minutes after the Big Bang, when the
universe was a few times hotter than the core of the sun. Third, the
cosmic microwave background (CMB), the afterglow of the Big Bang,
discovered in 1965 by Arno A. Penzias and Robert W. Wilson\cite{Wilson}
as a very isotropic blackbody radiation at a temperature of about 3
degrees Kelvin, emitted when the universe was cold enough to form
neutral atoms, and photons decoupled from matter, approximately 380,000
years after the Big Bang. Today, these observations are confirmed to
within a few percent accuracy, and have helped establish the hot Big
Bang as the preferred model of the universe.

Modern Cosmology begun as a quantitative science with the advent of
Einstein's general relativity and the realization that the
geometry of space-time, and thus the general attraction of matter, is
determined by the energy content of the universe,\cite{Weinberg}
\begin{equation}\label{EinsteinEquations}
G_{\mu\nu}\equiv R_{\mu\nu} - {1\over2}g_{\mu\nu}R +
\Lambda\,g_{\mu\nu} = 8\pi G\,T_{\mu\nu}
\,.
\end{equation}
These non-linear equations are simply too difficult to solve without
invoking some symmetries of the problem at hand: the universe itself.

We live on Earth, just 8 light-minutes away from our star, the Sun,
which is orbiting at 8.5 kpc from the center of our
galaxy,\footnote{One parallax second (1 pc), {\em parsec} for short,
corresponds to a distance of about 3.26 light-years or
$3.09\times10^{18}$ cm.} the Milky Way, an ordinary galaxy within the
Virgo cluster, of size a few Mpc, itself part of a supercluster of
size a few 100 Mpc, within the visible universe, approximately 10,000
Mpc in size. Although at small scales the universe looks very
inhomogeneous and anisotropic, the deepest galaxy catalogs like 2dF
GRS and SDSS suggest that the universe on large scales (beyond the
supercluster scales) is very homogeneous and isotropic. Moreover, the
cosmic microwave background, which contains information about the
early universe, indicates that the deviations from homogeneity and
isotropy were just a few parts per million at the time of photon
decoupling. Therefore, we can safely impose those symmetries to the
univerge at large and determine the corresponding evolution equations.
The most general metric satisfying homogeneity and isotropy is the
Friedmann-Robertson-Walker (FRW) metric, written here in terms of the
invariant geodesic distance $ds^2=g_{\mu\nu}dx^\mu dx^\nu$ in four
dimensions,\cite{Weinberg} $\mu=0,1,2,3$,\footnote{I am using $c=1$
everywhere, unless specified, and a metric signature $(-,+,+,+)$.}
\begin{equation}\label{FRWmetric}
ds^2 = - dt^2 + a^2(t)\left[{dr^2\over1-K\,r^2} +
r^2(d\theta^2 + \sin^2\theta\,d\phi^2)\right]\,,
\end{equation}
characterized by just two quantities, a {\em scale factor} $a(t)$,
which determines the physical size of the universe, and a constant $K$,
which characterizes the {\em spatial} curvature of the universe,
\begin{equation}\label{SpatialCurvature}
{}^{(3)}\!R = {6K\over a^2(t)} \hspace{2cm}
\left\{\begin{array}{ll}K=-1\hspace{1cm}&{\rm OPEN}\\
K=0&{\rm FLAT}\\K=+1&{\rm CLOSED}
\end{array}\right.
\end{equation}
Spatially open, flat and closed universes have different three-geometries.
Light geodesics on these universes behave differently, and thus could in
principle be distinguished observationally, as we shall discuss later.
Apart from the three-dimensional spatial curvature, we can also compute
a four-dimensional {\em space-time} curvature,
\begin{equation}\label{SpacetimeCurvature}
{}^{(4)}\!R = 6{\ddot a\over a} + 6\left({\dot a\over a}\right)^2 + 
6{K\over a^2}\,. 
\end{equation}
Depending on the dynamics (and thus on the matter/energy content) of the
universe, we will have different possible outcomes of its evolution.
The universe may expand for ever, recollapse in the future or approach
an asymptotic state in between.

\subsection{The matter and energy content of the universe}

The most general matter fluid consistent with the assumption of
homogeneity and isotropy is a perfect fluid, one in which an observer
{\em comoving with the fluid} would see the universe around it as
isotropic. The energy momentum tensor associated with such a fluid can
be written as\cite{Weinberg}
\begin{equation}\label{PerfectFluid}
T^{\mu\nu} = p\,g^{\mu\nu} + (p+\rho)\,U^\mu U^\nu\,,
\end{equation}
where $p(t)$ and $\rho(t)$ are the pressure and energy density of the
fluid at a given time in the expansion, as measured by this comoving
observer, and $U^\mu$ is the comoving four-velocity, satisfying $U^\mu
U_\mu=-1$. For such a comoving observer, the matter content looks
isotropic (in its rest frame),
\begin{equation}\label{PerfectFluidRestframe}
T^\mu_{\ \ \nu} = {\rm diag}(-\rho(t),\,p(t),\,p(t),\,p(t))\,.
\end{equation}
The conservation of energy ($T^{\mu\nu}_{\hspace{3mm};\nu} = 0$), a
direct consequence of the general covariance of the theory
($G^{\mu\nu}_{\hspace{3mm};\nu} = 0$), can be written in terms of the
FRW metric and the perfect fluid tensor (\ref{PerfectFluid}) as
\begin{equation}\label{EnergyConservation}
\dot\rho + 3{\dot a\over a}(p+\rho) = 0\,.
\end{equation}

In order to find explicit solutions, one has to supplement the
conservation equation with an {\em equation of state} 
relating the pressure and the density of the fluid, $p=p(\rho)$.
The most relevant fluids in cosmology are barotropic, i.e. fluids whose
pressure is linearly proportional to the density, $p=w\,\rho$, and
therefore the speed of sound is constant in those fluids.

We will restrict ourselves in these lectures to three main types of
barotropic fluids:

\begin{itemize}

\item {\em Radiation}, with equation of state $p_R=\rho_R/3$, associated with
relativistic degrees of freedom (i.e. particles with temperatures much
greater than their mass). In this case, the energy density of radiation
decays as $\rho_R \sim a^{-4}$ with the expansion of the universe.

\item {\em Matter}, with equation of state $p_M\simeq0$, associated with
nonrelativistic degrees of freedom (i.e. particles with temperatures
much smaller than their mass). In this case, the energy density of 
matter decays as $\rho_M \sim a^{-3}$ with the expansion of the universe.

\item {\em Vacuum energy}, with equation of state $p_V=-\rho_V$,
associated with quantum vacuum fluctuations. In this case, the vacuum
energy density remains constant with the expansion of the universe.

\end{itemize}

This is all we need in order to solve the Einstein equations.
Let us now write the equations of motion of observers comoving with
such a fluid in an expanding universe. According to general
relativity, these equations can be deduced from the Einstein equations
(\ref{EinsteinEquations}), by substituting the FRW metric
(\ref{FRWmetric}) and the perfect fluid tensor
(\ref{PerfectFluid}). The $\mu=i,\ \nu=j$ component of the Einstein
equations, together with the $\mu=0,\ \nu=0$ component constitute 
the so-called Friedmann equations,
\begin{eqnarray}
\label{FriedmannEquation}
\left({\dot a\over a}\right)^2 &=& {8\pi G\over3}\,\rho +
{\Lambda\over3} - {K\over a^2}\,,\\
{\ddot a\over a} &=& -\,{4\pi G\over3}\,(\rho + 3p) +
{\Lambda\over3} \,.\label{Evolution}
\end{eqnarray}
These equations contain all the relevant dynamics, since the energy
conservation equation (\ref{EnergyConservation}) can be obtained from these.

\subsection{The Cosmological Parameters}

I will now define the most important cosmological parameters. Perhaps
the best known is the {\em Hubble parameter} or rate of expansion today,
$H_0 = \dot a/a(t_0)$. We can write the Hubble parameter in units of
100 km\,s$^{-1}$Mpc$^{-1}$, which can be used to estimate the order of
magnitude for the present size and age of the universe,
\begin{eqnarray}
H_0&\equiv& 100\,h\ \ {\rm km\,s}^{-1}{\rm Mpc}^{-1}\,,\\
c\,H_0^{-1} &=& 3000\,h^{-1}\ {\rm Mpc}\,,\\
H_0^{-1} &=& 9.773\,h^{-1}\ {\rm Gyr}\,.
\end{eqnarray}
The parameter $h$ was measured to be in the range $0.4 < h< 1$ for
decades, and only in the last few years has it been found to lie within
4\% of $h=0.70$.  I will discuss those recent measurements in the next
Section.

Using the present rate of expansion, one can define a {\em critical}
density $\rho_c$, that which corresponds to a flat universe,
\begin{eqnarray}\label{CriticalDensity}
\rho_c\equiv{3H_0^2\over8\pi G}&=& 1.88\,h^2\,10^{-29}\ {\rm g/cm}^3\\
&=& 2.77\,h^{-1}\,10^{11}\ M_\odot/(h^{-1}\,{\rm Mpc})^3\\[1mm]
&=& 11.26\,h^2\ {\rm protons}/{\rm m}^3\,,
\end{eqnarray}
where $M_\odot=1.989\times10^{33}$ g \ is a solar mass unit. The
critical density $\rho_c$ corresponds to approximately 6 protons per
cubic meter, certainly a very dilute fluid! 

In terms of the critical density it is possible to define the 
{\rm density parameter}
\begin{equation}
\Omega_0\equiv{8\pi G\over3H_0^2}\,\rho(t_0)={\rho\over\rho_c}(t_0)\,,
\end{equation}
whose sign can be used to determine the spatial (three-)curvature.
Closed universes ($K=+1$) have $\Omega_0 > 1$, flat universes ($K=0$)
have $\Omega_0 = 1$, and open universes ($K=-1$) have $\Omega_0 < 1$,
no matter what are the individual components that sum up to the
density parameter.

In particular, we can define the individual ratios $\Omega_i \equiv
\rho_i/\rho_c$, for matter, radiation, cosmological constant and even
curvature, today, 
\begin{eqnarray}\label{Omega}
&&\Omega_M={8\pi G\,\rho_M\over3H_0^2}\hspace{2cm} 
\Omega_R={8\pi G\,\rho_R\over3H_0^2}\\
&&\,\Omega_\Lambda={\Lambda\over3H_0^2}\hspace{2.6cm}
\Omega_K=-\,{K\over a_0^2H_0^2}\,.
\end{eqnarray} 
For instance, we can evaluate today the radiation component
$\Omega_{\rm R}$, corresponding to relativistic particles, from the
density of microwave background photons, $\rho_{\rm CMB} =
\pi^2k^4T_{\rm CMB}^4/(15\hbar^3c^3) = 4.5\times 10^{-34}\
{\rm g/cm}^3$, which gives $\Omega_{\rm CMB} = 2.4\times 10^{-5}\
h^{-2}$. Three approximately massless neutrinos would contribute a
similar amount. Therefore, we can safely neglect the contribution of
relativistic particles to the total density of the universe today,
which is dominated either by non-relativistic particles (baryons, dark
matter or massive neutrinos) or by a cosmological constant, and write
the rate of expansion in terms of its value today, as
\begin{equation}\label{H2a}
H^2(a) = H_0^2\left(\Omega_R\,{a_0^4\over a^4} + 
\Omega_M\,{a_0^3\over a^3} + \Omega_\Lambda + 
\Omega_K\,{a_0^2\over a^2}\right)\,.
\end{equation}
An interesting consequence of these definitions is that one can now
write the Friedmann equation today, $a=a_0$, as a {\em cosmic sum rule}, 
\begin{equation}\label{CosmicSumRule}
1 = \Omega_M + \Omega_\Lambda + \Omega_K\,,
\end{equation}
where we have neglected $\Omega_{\rm R}$ today. That is, in the context
of a FRW universe, the total fraction of matter density, cosmological
constant and spatial curvature today must add up to one.  For instance,
if we measure one of the three components, say the spatial curvature, we
can deduce the sum of the other two. 

Looking now at the second Friedmann equation (\ref{Evolution}), we can
define another basic parameter, the {\em deceleration parameter},
\begin{equation}\label{Deceleration}
q_0 = - {a\,\ddot a\over\dot a^2}(t_0) = {4\pi G\over3H_0^2}\,
\Big[\rho(t_0)+3p(t_0)\Big]\,,
\end{equation}
defined so that it is positive for ordinary matter and radiation,
expressing the fact that the universe expansion should slow down due
to the gravitational attraction of matter. We can write this parameter
using the definitions of the density parameter for known and unknown
fluids (with density $\Omega_x$ and arbitrary equation of state $w_x$) as
\begin{equation}\label{DecelerationParameter}
q_0 = \Omega_R + \half\Omega_M - \Omega_\Lambda + \half
\sum_x (1+3w_x)\,\Omega_x\,.
\end{equation}
Uniform expansion corresponds to $q_0=0$ and requires a cancellation
between the matter and vacuum energies. For matter domination, $q_0 >0$,
while for vacuum domination, $q_0 < 0$. As we will see in a moment,
we are at present probing the time dependence of the deceleration
parameter and can determine with some accuracy the moment at which
the universe went from a decelerating phase, dominated by dark matter,
into an acceleration phase at present, which seems to indicate the
dominance of some kind of vacuum energy.

\subsection{The $(\Omega_M,\,\Omega_\Lambda)$ plane}

Now that we know that the universe is accelerating, one can parametrize
the matter/energy content of the universe with just two components:
the matter, characterized by $\Omega_M$, and the vacuum energy 
$\Omega_\Lambda$. Different values of these two parameters completely
specify the universe evolution. It is thus natural to plot the results
of observations in the plane ($\Omega_M,\ \Omega_\Lambda$), in order to
check whether we arrive at a consistent picture of the present universe
from several different angles (different sets of cosmological observations).

\begin{figure}[t]
\epsfxsize=9cm 
\epsfbox{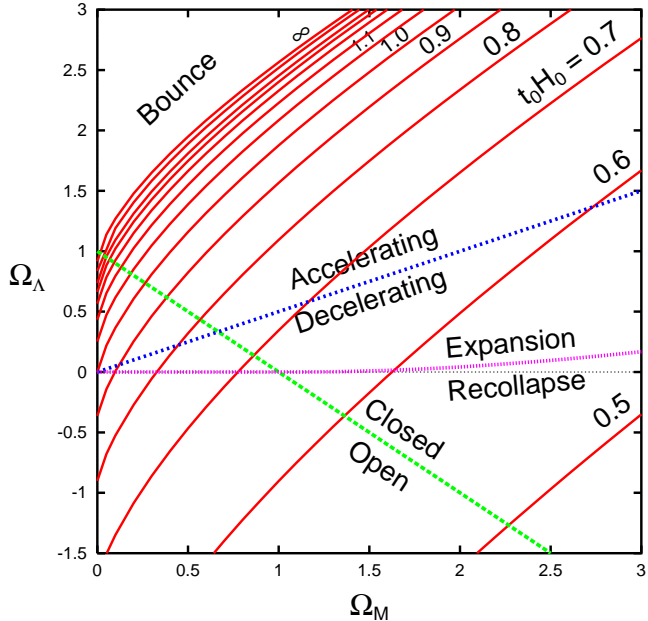} 
\vspace*{-1cm}
\caption{Parameter space $(\Omega_M,\,\Omega_\Lambda)$. The green
(dashed) line $\Omega_\Lambda=1-\Omega_M$ corresponds to a flat
universe, $\Omega_K=0$, separating open from closed universes. The blue
(dotted) line $\Omega_\Lambda=\Omega_M/2$ corresponds to uniform
expansion, $q_0 = 0$, separating accelerating from decelerating
universes. The violet (dot-dashed) line corresponds to critical
universes, separating eternal expansion from recollapse in the
future. Finally, the red (continuous) lines correspond to
$t_0H_0=0.5,\,0.6,\dots,\,\infty$, beyond which the universe has a
bounce.
\label{fig:OMOLplane}}
\end{figure}

Moreover, different regions of this plane specify different behaviors
of the universe. The boundaries between regions are well defined curves
that can be computed for a given model. I will now describe the various
regions and boundaries.

\begin{itemize}

\item {\em Uniform expansion} ($q_0=0$). Corresponds to the line
$\Omega_\Lambda=\Omega_M/2$. Points above this line correspond to
universes that are accelerating today, while those below correspond to
decelerating universes, in particular the old cosmological model of
Einstein-de Sitter (EdS), with $\Omega_\Lambda=0,\ \Omega_M=1$.  Since
1998, all the data from Supernovae of type Ia appear above this line, 
many standard deviations away from EdS universes.

\item {\em Flat universe} ($\Omega_K=0$). Corresponds to the line
$\Omega_\Lambda=1-\Omega_M$. Points to the right of this line
correspond to closed universes, while those to the left correspond to
open ones.  In the last few years we have mounting evidence that the
universe is spatially flat (in fact Euclidean).

\item {\em Bounce} ($t_0H_0=\infty$). Corresponds to a complicated
function of $\Omega_\Lambda(\Omega_M)$, normally expressed as an
integral equation, where
$$t_0H_0=\int_0^1 da\ [1+\Omega_M(1/a-1)+\Omega_\Lambda(a^2-1)]^{-1/2}$$
is the product of the age of the universe and the present rate of
expansion. Points above this line correspond to universes that have
contracted in the past and have later rebounced. At present, these 
universes are ruled out by observations of galaxies and quasars at
high redshift (up to $z=10$).

\item {\em Critical Universe} ($H=\dot H=0$). Corresponds to the
boundary between eternal expansion in the future and recollapse. For
$\Omega_M \leq1$, it is simply the line $\Omega_\Lambda=0$, but for
$\Omega_M >1$, it is a more complicated curve,
$$\Omega_\Lambda=4\Omega_M\sin^3\Big[{1\over3}\arcsin\Big({\Omega_M-
1\over\Omega_M}\Big)\Big]\simeq{4\over27}{(\Omega_M-1)^3\over\Omega_M^2}.$$
These critical solutions are asymptotic to the EdS model.

\end{itemize}

These boundaries, and the regions they delimit,
can be seen in Fig.~1, together with the lines of equal 
$t_0H_0$ values.

In summary, the basic cosmological parameters that are now been hunted 
by a host of cosmological observations are the following: the present
rate of expansion $H_0$; the age of the universe $t_0$; the
deceleration parameter $q_0$; the spatial curvature $\Omega_K$; the
matter content $\Omega_M$; the vacuum energy $\Omega_\Lambda$; the
baryon density $\Omega_B$; the neutrino density $\Omega_\nu$, and
many other that characterize the perturbations responsible for the
large scale structure (LSS) and the CMB anisotropies.

\section{The accelerating universe}

Let us first describe the effect that the expansion of the universe
has on the objects that live in it. In the absence of other forces
but those of gravity, the trajectory of a particle is given by
general relativity in terms of the geodesic equation
\begin{equation}\label{geodesic}
{du^\mu\over ds} + \Gamma^\mu_{\nu\lambda}\,u^\nu u^\lambda =0\,,
\end{equation}
where $u^\mu=(\gamma,\,\gamma v^i)$, with $\gamma^2=1-v^2$ and $v^i$ is
the peculiar velocity. Here $\Gamma^\mu_{\nu\lambda}$ is the Christoffel
connection,\cite{Weinberg} whose only non-zero component is
$\Gamma^0_{ij} = (\dot a/a)\,g_{ij}$; substituting into the geodesic
equation, we obtain $|\vec u|\propto 1/a$, and thus the particle's
momentum decays with the expansion like $p\propto 1/a$. In the case of a
photon, satisfying the de Broglie relation $p=h/\lambda$, one obtains
the well known {\em photon redshift}
\begin{equation}\label{redshift}
{\lambda_1\over\lambda_0} = {a(t_1)\over a(t_0)} \ \ \Rightarrow \ 
\ z\equiv{\lambda_0-\lambda_1\over\lambda_1} = {a_0\over a_1} - 1\,,
\end{equation}
where $\lambda_0$ is the wavelength measured by an observer at time
$t_0$, while $\lambda_1$ is the wavelength emitted when the universe was
younger $(t_1<t_0)$. Normally we measure light from stars in distant
galaxies and compare their observed spectra with our laboratory
(restframe) spectra. The fraction (\ref{redshift}) then gives the redshift
$z$ of the object. We are assuming, of course, that both the emitted and
the restframe spectra are identical, so that we can actually measure the
effect of the intervening expansion, i.e. the growth of the scale factor
from $t_1$ to $t_0$, when we compare the two spectra. Note that if the
emitting galaxy and our own participated in the expansion, i.e. if our
measuring rods (our rulers) also expanded with the universe, we would
see no effect!  The reason we can measure the redshift of light from a
distant galaxy is because our galaxy is a gravitationally bounded
object that has decoupled from the expansion of the universe. It is the
distance between galaxies that changes with time, not the sizes of
galaxies, nor the local measuring rods.

We can now evaluate the relationship between physical distance and
redshift as a function of the rate of expansion of the universe.
Because of homogeneity we can always choose our position to be at the
origin $r=0$ of our spatial section. Imagine an object (a star)
emitting light at time $t_1$, at coordinate distance $r_1$ from the
origin. Because of isotropy we can ignore the angular coordinates
$(\theta,\phi)$. Then the physical distance, to first order, will
be $d=a_0\,r_1$. Since light travels along null geodesics,\cite{Weinberg}
we can write $0 = -dt^2+a^2(t)\,dr^2/(1-Kr^2)$, and therefore,
\begin{equation}\label{distance}
\int_{t_1}^{t_0}{dt\over a(t)} = \int_0^{r_1}{dr\over\sqrt{1-Kr^2}} 
\equiv f(r_1) = \left\{\begin{array}{ll}\arcsin r_1 \hspace{1cm} & K=1\\
r_1 & K=0\\{\rm arcsinh}\, r_1 & K=-1\end{array}\right.
\end{equation}
If we now Taylor expand the scale factor to first order,
\begin{equation}\label{Taylor}
{1\over1+z} = {a(t)\over a_0} = 1 + H_0(t-t_0)+{\cal O}(t-t_0)^2\,,
\end{equation}
we find, to first approximation,
$$r_1 \approx f(r_1) = {1\over a_0}(t_0 - t_1) + \dots = 
{z\over a_0H_0} + \dots$$
Putting all together we find the famous Hubble law
\begin{equation}\label{HubbleLaw}
H_0\,d = a_0H_0r_1 = z \simeq vc\,,
\end{equation}
which is just a kinematical effect (we have not included yet
any dynamics, i.e. the matter content of the universe). Note that at
low redshift $(z\ll1)$, one is tempted to associate the observed
change in wavelength with a Doppler effect due to a hypothetical
recession velocity of the distant galaxy. This is only an
approximation. In fact, the redshift cannot be ascribed to the
relative velocity of the distant galaxy because in general relativity
(i.e. in curved spacetimes) one cannot compare velocities through
parallel transport, since the value depends on the path! If the
distance to the galaxy is small, i.e. $z\ll1$, the physical spacetime
is not very different from Minkowsky and such a comparison is
approximately valid. As $z$ becomes of order one, such a relation is
manifestly false: galaxies cannot travel at speeds greater than the
speed of light; it is the stretching of spacetime which is responsible
for the observed redshift.

\begin{figure}[t]
\epsfxsize=6cm 
\epsfbox{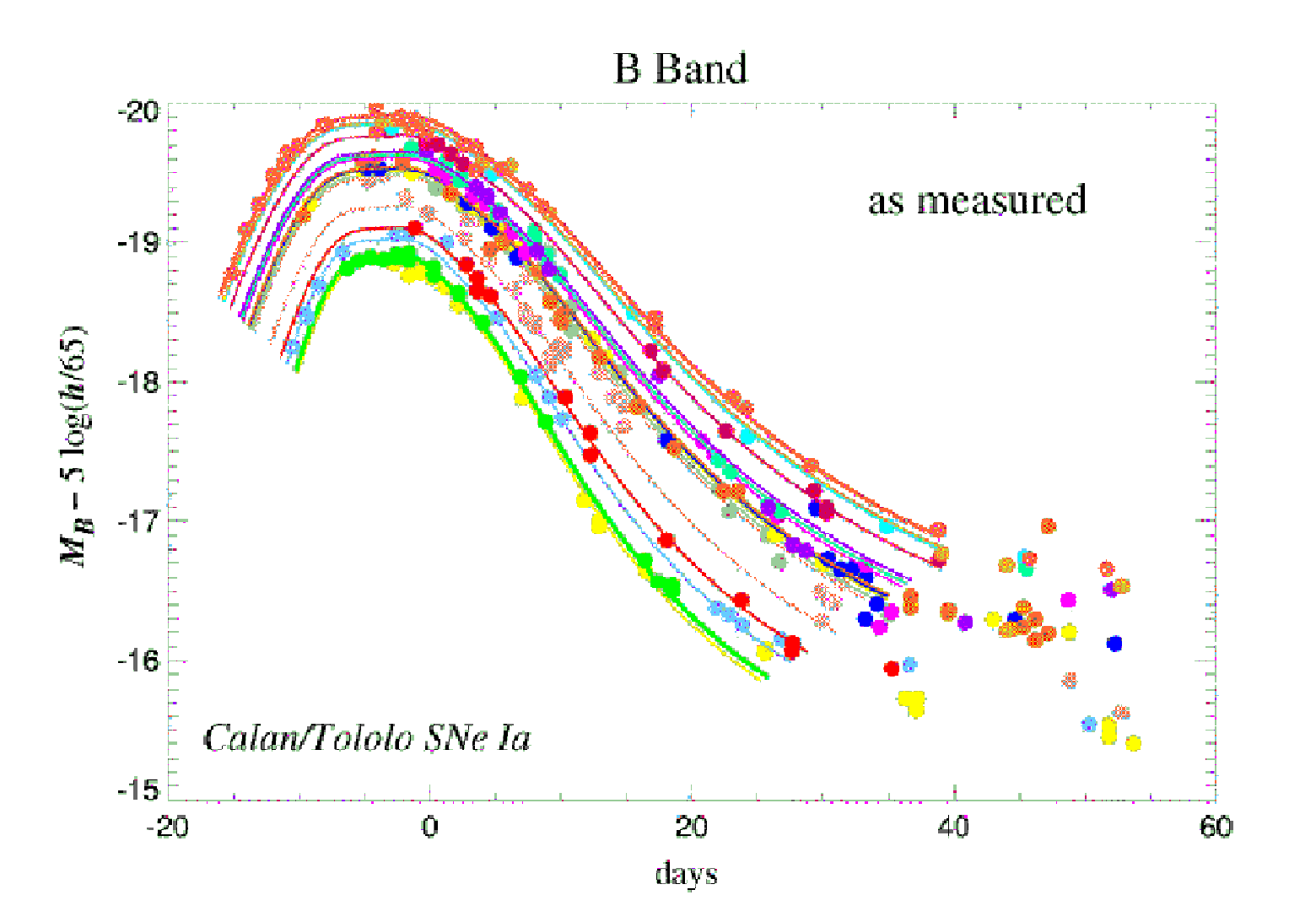} 
\epsfxsize=6cm 
\epsfbox{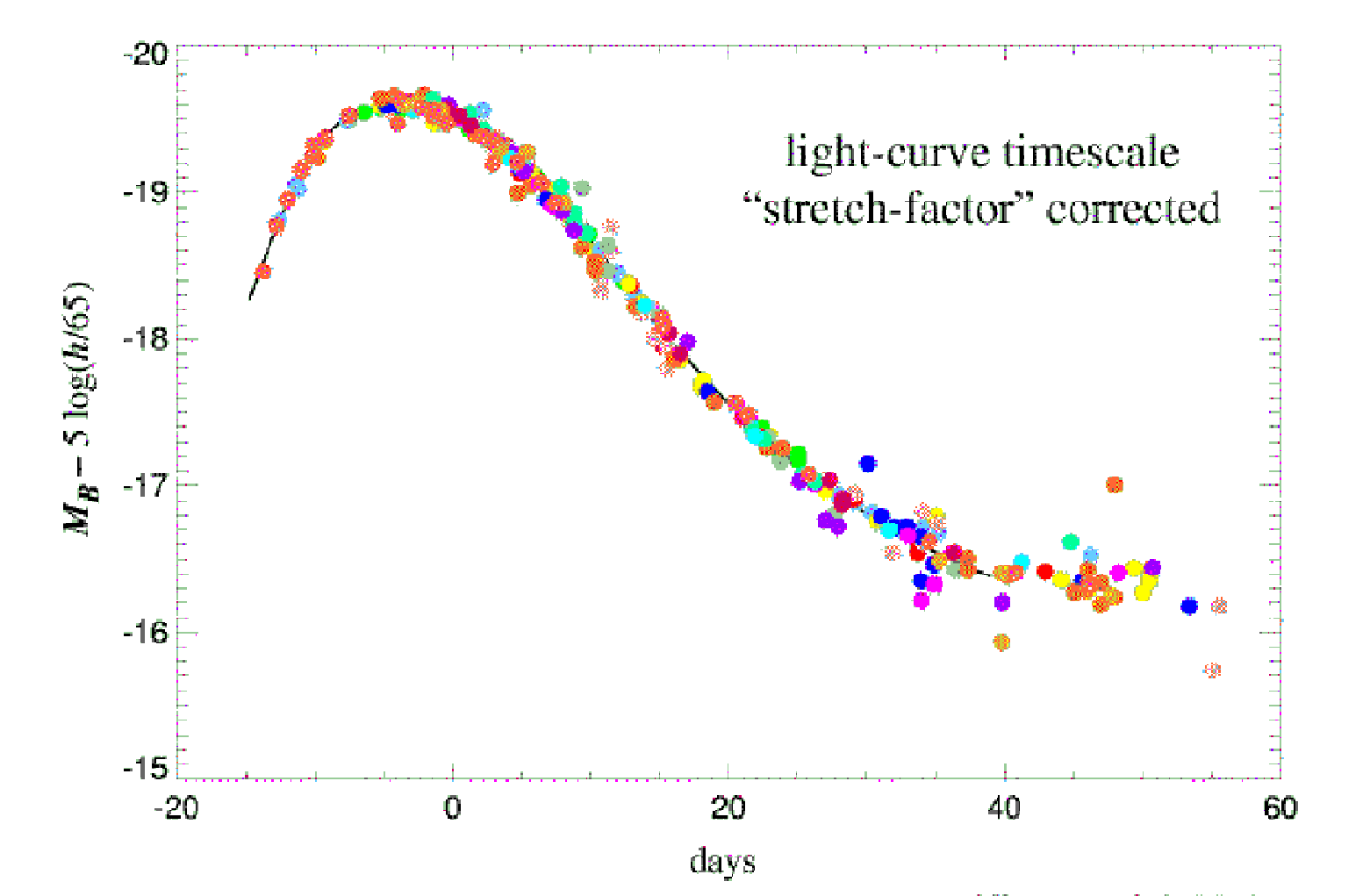} 
\caption{The Type Ia supernovae observed nearby show a relationship
between their absolute luminosity and the timescale of their light
curve: the brighter supernovae are slower and the fainter ones are
faster. A simple linear relation between the absolute magnitude and a
``stretch factor'' multiplying the light curve timescale fits the data
quite well.
\label{fig:stretchfactor}}
\end{figure}

Hubble's law has been confirmed by observations ever since the 1920s,
with increasing precision, which have allowed cosmologists to determine
the Hubble parameter $H_0$ with less and less systematic errors. 
Nowadays, the best determination of the Hubble parameter was made by
the Hubble Space Telescope Key Project,\cite{HSTKP} 
$H_0 = 72 \pm 8$ km/s/Mpc. This determination is based on objects
at distances up to 500 Mpc, corresponding to redshifts $z\leq0.1$.

Nowadays, we are beginning to probe much greater distances,
corresponding to $z\simeq1$, thanks to type Ia supernovae.  These are
white dwarf stars at the end of their life cycle that accrete matter
from a companion until they become unstable and violently explode in a
natural thermonuclear explosion that out-shines their progenitor
galaxy. The intensity of the distant flash varies in time, it takes
about three weeks to reach its maximum brightness and then it declines
over a period of months.  Although the maximum luminosity varies from
one supernova to another, depending on their original mass, their
environment, etc., there is a pattern: brighter explosions last longer
than fainter ones. By studying the characteristic light curves, see
Fig.~2, of a reasonably large statistical sample, cosmologists from
the Supernova Cosmology Project\cite{SCP} and the High-redshift
Supernova Project,\cite{HRS} are now quite confident that they can
use this type of supernova as a standard candle. Since the light
coming from some of these rare explosions has travelled a large
fraction of the size of the universe, one expects to be able to infer
from their distribution the spatial curvature and the rate of
expansion of the universe. The connection between observations of high
redshift supernovae and cosmological parameters is done via the
luminosity distance, defined as the distance $d_L$ at which a source
of absolute luminosity (energy emitted per unit time) ${\cal L}$ gives
a flux (measured energy per unit time and unit area of the detector)
${\cal F}={\cal L}/4\pi\,d_L^2$. One can then evaluate, within a given
cosmological model, the expression for $d_L$ as a function of
redshift,\cite{JGB}
\begin{equation}\label{LuminosityDistance}
H_0\,d_L(z) = {(1+z)\over|\Omega_K|^{1/2}}\,{\rm sinn}\left[
\int_0^z {|\Omega_K|^{1/2}\ dz'\over\sqrt{(1+z')^2(1+z'\Omega_M) - 
z'(2+z')\Omega_\Lambda}}\right]\,,
\end{equation}
where ${\rm sinn}(x)= x\ {\rm if}\ K=0;\ \sin(x)\ {\rm if}\ K=+1\ {\rm
and}\ \sinh(x)\ {\rm if}\ K=-1$, and we have used the cosmic sum rule
(\ref{CosmicSumRule}). 

\begin{figure}[t]
\epsfysize=9cm 
\epsfbox{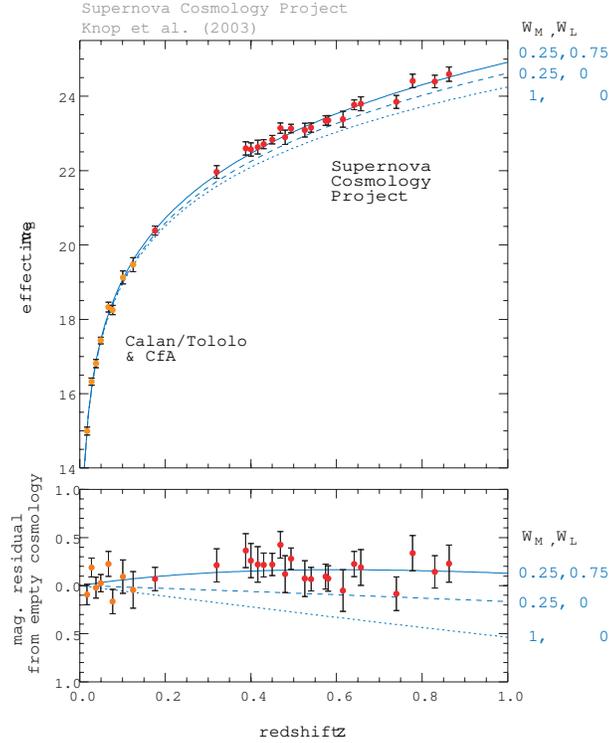} 
\caption{Upper panel: The Hubble diagram in linear redshift
scale. Supernovae with $\Delta z < 0.01$ of eachother have been
weighted-averaged binned.  The solid curve represents the best-fit flat
universe model, $(\Omega_M=0.25,\,\Omega_\Lambda=0.75)$. Two other
cosmological models are shown for comparison,
$(\Omega_M=0.25,\,\Omega_\Lambda=0)$ and
$(\Omega_M=1,\,\Omega_\Lambda=0)$. Lower panel: Residuals of the
averaged data relative to an empty universe.
\label{fig:Knop03}}
\end{figure}

Astronomers measure the relative luminosity of a distant object in
terms of what they call the effective magnitude, which has a peculiar
relation with distance,
\begin{equation}\label{EffectiveMagnitude}
m(z) \equiv M + 5\,\log_{10}\Big[{d_L(z)\over{\rm Mpc}}\Big] + 25
= \bar M + 5\,\log_{10}[H_0\,d_L(z)]\,.
\end{equation}
Since 1998, several groups have obtained serious evidence that high
redshift supernovae appear fainter than expected for either an open
$(\Omega_M < 1)$ or a flat $(\Omega_M = 1)$ universe, see Fig.~3.  In
fact, the universe appears to be accelerating instead of decelerating,
as was expected from the general attraction of matter, see
Eq.~(\ref{DecelerationParameter}); something seems to be acting as a
repulsive force on very large scales. The most natural explanation for
this is the presence of a cosmological constant, a diffuse vacuum
energy that permeates all space and, as explained above, gives the
universe an acceleration that tends to separate gravitationally bound
systems from each other. The best-fit results from the Supernova
Cosmology Project\cite{SCP2003} give a linear combination
$$0.8\Omega_M - 0.6\Omega_\Lambda = - 0.16 \pm 0.05 \ \ (1\sigma),$$
which is now many sigma away from an EdS model with $\Lambda=0$.
In particular, for a flat universe this gives 
$$\Omega_\Lambda = 0.71 \pm 0.05 \hspace{1cm} {\rm and} \hspace{1cm} 
\Omega_M = 0.29 \pm 0.05 \ \ (1\sigma).$$ 
Surprising as it may seem, arguments for a significant dark energy
component of the universe where proposed long before these
observations, in order to accommodate the ages of globular clusters,
as well as a flat universe with a matter content below critical, which
was needed in order to explain the observed distribution of galaxies,
clusters and voids.

Taylor expanding the scale factor to third order,
\begin{equation}
{a(t)\over a_0} = 1 + H_0(t-t_0) - {q_0\over2!}H_0^2(t-t_0)^2 + 
{j_0\over3!}H_0^3(t-t_0)^3 + {\cal O}(t-t_0)^4\,,
\end{equation}
where
\begin{eqnarray}
&&q_0=-\,{\ddot a\over a H^2}(t_0) = \half\sum_i(1+3w_i)\Omega_i=
\half\Omega_M-\Omega_\Lambda\,, \\
&&j_0=+\,{\stackrel{\dots}{a}\over a H^3}(t_0) = \half\sum_i(1+3w_i)
(2+3w_i)\Omega_i=\Omega_M+\Omega_\Lambda\,,\label{JerkParameter}
\end{eqnarray}
are the deceleration and ``jerk'' parameters. Substituting into
Eq.~(\ref{LuminosityDistance}) we find
\begin{equation}\label{DLZ}
H_0\,d_L(z) = z + {1\over2}(1-q_0)\,z^2 - 
{1\over6}(1-q_0-3q_0^2+j_0)\,z^3 + {\cal O}(z^4)\,.
\end{equation}
This expression goes beyond the leading linear term, corresponding to
the Hubble law, into the second and third order terms, which are
sensitive to the cosmological parameters $\Omega_M$ and
$\Omega_\Lambda$. It is only recently that cosmological observations
have gone far enough back into the early universe that we can begin to
probe these terms, see Fig.~4.

\begin{figure}[t]
\epsfxsize=8cm 
\epsfbox{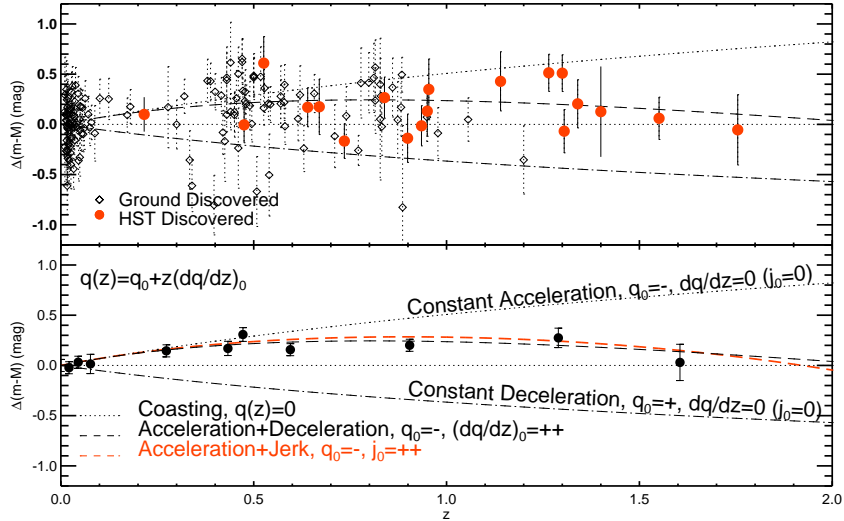} 
\vspace*{-3.3cm}
\caption{The Supernovae Ia residual Hubble diagram. Upper panel:
Ground-based discoveries are represented by diamonds, HST-discovered
SNe Ia are shown as filled circles. Lower panel: The same but with
weighted averaged in fixed redshift bins. Kinematic models of the
expansion history are shown relative to an eternally coasting model
$q(z)=0$.
\label{fig:Riess04}}
\end{figure}

This extra component of the critical density would have to resist
gravitational collapse, otherwise it would have been detected already
as part of the energy in the halos of galaxies.  However, if most of
the energy of the universe resists gravitational collapse, it is
impossible for structure in the universe to grow. This dilemma can be
resolved if the hypothetical dark energy was negligible in the past
and only recently became the dominant component. According to general
relativity, this requires that the dark energy have {\rm negative}
pressure, since the ratio of dark energy to matter density goes like
$a(t)^{-3p/\rho}$. This argument would rule out almost all of the
usual suspects, such as cold dark matter, neutrinos, radiation, and
kinetic energy, since they all have zero or positive pressure. Thus,
we expect something like a cosmological constant, with a negative
pressure, $p\approx-\rho$, to account for the missing energy.

However, if the universe was dominated by dark matter in the past, in
order to form structure, and only recently became dominated by dark
energy, we must be able to see the effects of the transition from the
deceleration into the acceleration phase in the luminosity of distant
type Ia supernovae. This has been searched for since 1998, when the
first convincing results on the present acceleration appeared.
However, only recently\cite{Riess2004} do we have clear evidence of
this transition point in the evolution of the universe.  This {\em
coasting point} is defined as the time, or redshift, at which the
deceleration parameter vanishes,
\begin{equation}
q(z) = -1 + (1+z)\,{d\over dz}\ln H(z) = 0\,,
\end{equation}
where
\begin{equation}
H(z) = H_0\Big[\Omega_M (1+z)^3 + \Omega_x\,e^{3\int_0^z(1+w_x(z'))
{dz'\over1+z'}}\ + \Omega_K (1+z)^2\Big]^{1/2}\,,
\end{equation}
and we have assumed that the dark energy is parametrized by a
density $\Omega_x$ today, with a redshift-dependent equation of 
state, $w_x(z)$, not necessarily equal to $-1$. Of course, in the
case of a true cosmological constant, this reduces to the usual 
expression. 

\begin{figure}[t]
\epsfxsize=8cm 
\epsfbox{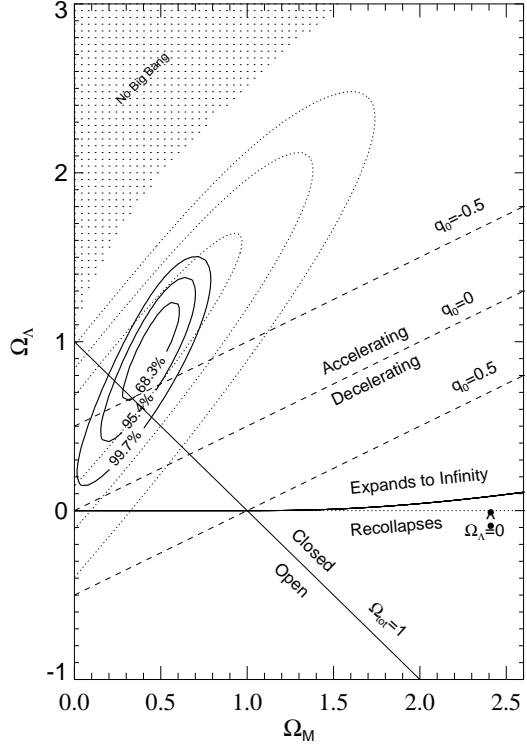} 
\vspace*{5mm}
\caption{The recent supernovae data on the
$(\Omega_M,\,\Omega_\Lambda)$ plane, see Ref. [12]. Shown are the 
1-, 2- and 3-$\sigma$ contours, as well as the data from 1998,
for comparison. It is clear that the old EdS cosmological model at 
$(\Omega_M=1,\,\Omega_\Lambda=0)$ is many standard deviations away
from the data. \label{fig:RiessOMOL}}
\end{figure}

Let us suppose for a moment that the barotropic parameter $w$ is
constant, then the coasting redshift can be determined from
\begin{eqnarray}
&&q(z) = \half\Big[{\Omega_M + (1+3w)\,\Omega_x\,(1+z)^{3w}\over
\Omega_M + \Omega_x\,(1+z)^{3w} + \Omega_K (1+z)^{-1}}\Big] = 0\,,\\
&&\hspace{1cm}\Rightarrow z_c = \left({(3|w|-1)\Omega_x\over\Omega_M}
\right)^{1\over3|w|}-1\,,
\label{DecelerationParamz}
\end{eqnarray}
which, in the case of a true cosmological constant, reduces to
\begin{equation}\label{CoastingPoint}
z_c = \Big({2\Omega_\Lambda\over\Omega_M}\Big)^{1/3}-1\,.
\end{equation}
When substituting $\Omega_\Lambda\simeq0.7$ and $\Omega_M\simeq0.3$,
one obtains $z_c \simeq 0.6$, in excellent agreement with recent
observations.\cite{Riess2004} The plane $(\Omega_M,\,\Omega_\Lambda)$
can be seen in Fig.~5, which shows a significant improvement with
respect to previous data.

Now, if we have to live with this vacuum energy, we might as well try
to comprehend its origin. For the moment it is a complete mystery,
perhaps the biggest mystery we have in physics today. We measure its
value but we don't understand why it has the value it has. In fact, if
we naively predict it using the rules of quantum mechanics, we find a
number that is many (many!) orders of magnitude off the mark. Let us
describe this calculation in some detail. In non-gravitational
physics, the zero-point energy of the system is irrelevant because
forces arise from gradients of potential energies. However, we know
from general relativity that even a constant energy density
gravitates.  Let us write down the most general energy momentum tensor
compatible with the symmetries of the metric and that is covariantly
conserved.  This is precisely of the form $T_{\mu\nu}^{(vac)} =
p_V\,g_{\mu\nu} = -\,\rho_V\,g_{\mu\nu}$. Substituting into the
Einstein equations (\ref{EinsteinEquations}), we see that the
cosmological constant and the vacuum energy are completely equivalent,
$\Lambda = 8\pi G\,\rho_V$, so we can measure the vacuum energy
with the observations of the acceleration of the universe, which
tells us that $\Omega_\Lambda\simeq0.7$.

On the other hand, we can estimate the contribution to the vacuum
energy coming from the quantum mechanical zero-point energy of the
quantum oscillators associated with the fluctuations of all quantum
fields,
\begin{equation}\label{ZeroPointEnergy}
\rho_V^{th} = \sum_i \int_0^{\Lambda_{UV}}\!\!{d^2k\over(2\pi)^3}\,
\half\hbar\omega_i(k)={\hbar\Lambda_{UV}^4\over16\pi^2}\sum_i
(-1)^F N_i + {\cal O}(m_i^2\Lambda_{UV}^2)\,,
\end{equation}
where $\Lambda_{UV}$ is the ultraviolet cutoff signaling the scale of
new physics. Taking the scale of quantum gravity, $\Lambda_{UV} =
M_{Pl}$, as the cutoff, and barring any fortuituous cancellations,
then the theoretical expectation (\ref{ZeroPointEnergy}) appears to be
120 orders of magnitude larger than the observed vacuum energy
associated with the acceleration of the universe,
\begin{eqnarray}\label{vacuumenergyth}
&&\rho_V^{th} \simeq 1.4\times 10^{74} \ {\rm GeV}^4  =
3.2\times 10^{91} \ {\rm g/cm}^3\,,\\
&&\rho_V^{obs} \simeq 0.7\,\rho_c = 0.66\times 10^{-29} \ {\rm g/cm}^3 =
2.9\times 10^{-11} \ {\rm eV}^4\,.\label{vacuumenergyobs}
\end{eqnarray}
Even if we assumed that the ultraviolet cutoff associated with quantum
gravity was as low as the electroweak scale (and thus around the
corner, liable to be explored in the LHC), the theoretical expectation
would still be 60 orders of magnitude too bit. This is by far the
worst mismatch between theory and observations in all of science.
There must be something seriously wrong in our present understanding
of gravity at the most fundamental level. Perhaps we don't understand
the vacuum and its energy does not gravitate after all, or perhaps we
need to impose a new principle (or a symmetry) at the quantum gravity
level to accommodate such a flagrant mismatch. 

In the meantime, one can at least parametrize our ignorance by making
variations on the idea of a {\em constant} vacuum energy. Let us
assume that it actually evolves slowly with time. In that case, we do
not expect the equation of state $p=-\rho$ to remain true, but instead
we expect the barotropic parameter $w(z)$ to depend on redshift.  Such
phenomenological models have been proposed, and until recently
produced results that were compatible with $w=-1$ today, but with
enough uncertainty to speculate on alternatives to a truly constant
vacuum energy.  However, with the recent supernovae
results,\cite{Riess2004} there seems to be little space for
variations, and models of a time-dependent vacuum energy are less and
less favoured. In the near future, the SNAP satellite\cite{SNAP} will
measure several thousand supernovae at high redshift and therefore map
the redshift dependence of both the dark energy density and its
equation of state with great precision. This will allow a much better
determination of the cosmological parameters $\Omega_M$ and
$\Omega_\Lambda$.

\section{Dark Matter}

In the 1920s Hubble realized that the so called nebulae were actually
distant galaxies very similar to our own. Soon afterwards, in 1933,
Zwicky found dynamical evidence that there is possibly ten to a hundred
times more mass in the Coma cluster than contributed by the luminous
matter in galaxies.\cite{Zwicky} However, it was not until the 1970s
that the existence of dark matter began to be taken more seriously. At
that time there was evidence that rotation curves of galaxies did not
fall off with radius and that the dynamical mass was increasing with
scale from that of individual galaxies up to clusters of galaxies. Since
then, new possible extra sources to the matter content of the universe
have been accumulating:
\begin{eqnarray}\label{MatterContent}
\Omega_M &=& \Omega_{B,\ {\rm lum}} \hspace{0.9cm} 
({\rm stars\ in\ galaxies})\\
&+& \Omega_{B,\ {\rm dark}} \hspace{0.8cm} ({\rm MACHOs?})\\
&+& \Omega_{CDM} \hspace{1cm} ({\rm weakly\ interacting:\ 
axion,\ neutralino?})\\
&+& \Omega_{HDM} \hspace{1cm} ({\rm massive\ neutrinos?})
\end{eqnarray}

The empirical route to the determination of $\Omega_M$ is nowadays one
of the most diversified of all cosmological parameters. The matter
content of the universe can be deduced from the mass-to-light ratio of
various objects in the universe; from the rotation curves of galaxies;
from microlensing and the direct search of Massive Compact Halo Objects
(MACHOs); from the cluster velocity dispersion with the use of the Virial
theorem; from the baryon fraction in the X-ray gas of clusters; from
weak gravitational lensing; from the observed matter distribution of the
universe via its power spectrum; from the cluster abundance and its
evolution; from direct detection of massive neutrinos at
SuperKamiokande; from direct detection of Weakly Interacting Massive
Particles (WIMPs) at DAMA and UKDMC, and finally from microwave background
anisotropies. I will review here just a few of them.

\subsection{Rotation curves of spiral galaxies}

The flat rotation curves of spiral galaxies provide the most direct
evidence for the existence of large amounts of dark matter. Spiral
galaxies consist of a central bulge and a very thin disk, stabilized
against gravitational collapse by angular momentum conservation, and
surrounded by an approximately spherical halo of dark matter. One can
measure the orbital velocities of objects orbiting around the disk as
a function of radius from the Doppler shifts of their spectral lines. 

\begin{figure}[t]
\begin{center}
\includegraphics[width=7cm,angle=0]{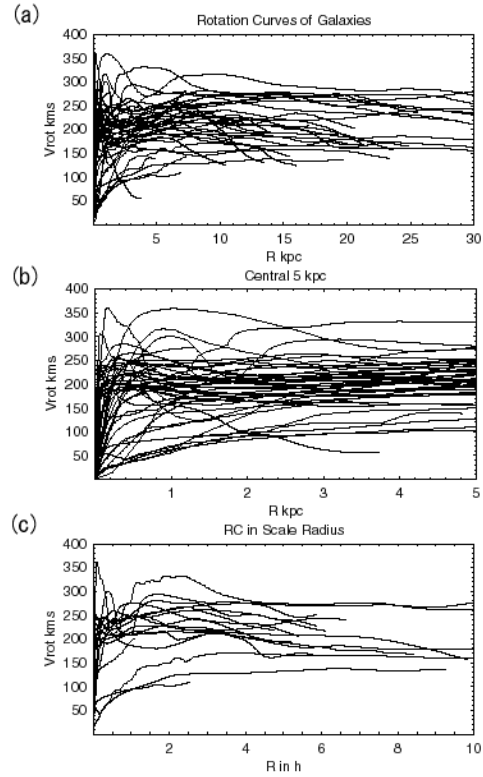}
\end{center}
\caption{The rotation curves of several hundred galaxies. Upper panel:
As a function of their radii in kpc. Middle panel: The central 5 kpc.
Lower panel: As a function of scale radius.}
\label{fig:NGC6503}
\end{figure}

The rotation curve of the Andromeda galaxy was first measured by
Babcock in 1938, from the stars in the disk. Later it became possible
to measure galactic rotation curves far out into the disk, and a trend
was found.\cite{Freeman} The orbital velocity rose linearly from the
center outward until it reached a typical value of 200 km/s, and then
remained flat out to the largest measured radii. This was completely
unexpected since the observed surface luminosity of the disk falls off
exponentially with radius,\cite{Freeman} $I(r) = I_0
\exp(-r/r_D)$. Therefore, one would expect that most of the galactic
mass is concentrated within a few disk lengths $r_D$, such that the
rotation velocity is determined as in a Keplerian orbit, $v_{\rm rot}
= (GM/r)^{1/2} \propto r^{-1/2}$. No such behaviour is observed. In
fact, the most convincing observations come from radio emission (from
the 21 cm line) of neutral hydrogen in the disk, which has been
measured to much larger galactic radii than optical tracers. A typical
case is that of the spiral galaxy NGC 6503, where $r_D = 1.73$ kpc,
while the furthest measured hydrogen line is at $r=22.22$ kpc, about
13 disk lengths away.  Nowadays, thousands of galactic rotation curves
are known, see Fig.~6, and all suggest the existence of about ten
times more mass in the halos of spiral galaxies than in the stars of
the disk. Recent numerical simulations of galaxy formation in a CDM
cosmology\cite{Frenk} suggest that galaxies probably formed by the
infall of material in an overdense region of the universe that had
decoupled from the overall expansion.  The dark matter is supposed to
undergo violent relaxation and create a virialized system, i.e. in
hydrostatic equilibrium. This picture has led to a simple model of
dark-matter halos as isothermal spheres, with density profile $\rho(r)
= \rho_c/(r_c^2 + r^2)$, where $r_c$ is a core radius and $\rho_c =
v_\infty^2/4\pi G$, with $v_\infty$ equal to the plateau value of the
flat rotation curve. This model is consistent with the universal
rotation curves seen in Fig.~6. At large radii the dark matter
distribution leads to a flat rotation curve. The question is for how
long. In dense galaxy clusters one expects the galactic halos to
overlap and form a continuum, and therefore the rotation curves should
remain flat from one galaxy to another. However, in field galaxies,
far from clusters, one can study the rotation velocities of
substructures (like satellite dwarf galaxies) around a given galaxy,
and determine whether they fall off at sufficiently large distances
according to Kepler's law, as one would expect, once the edges of the
dark matter halo have been reached. These observations are rather
difficult because of uncertainties in distinguishing between true
satellites and interlopers.  Recently, a group from the Sloan Digital
Sky Survey Collaboration claim that they have seen the edges of the
dark matter halos around field galaxies by confirming the fall-off at
large distances of their rotation curves.\cite{Klypin} These results, if
corroborated by further analysis, would constitute a tremendous
support to the idea of dark matter as a fluid surrounding galaxies and
clusters, while at the same time eliminates the need for modifications
of Newtonian of even Einstenian gravity at the scales of galaxies, to
account for the flat rotation curves.

That's fine, but how much dark matter is there at the galactic scale?
Adding up all the matter in galactic halos up to a maximum radii, one 
finds
\begin{equation}\label{OmegaHalo}
\Omega_{\rm halo} \simeq 10\ \Omega_{\rm lum} \geq 0.03 - 0.05\,.
\end{equation}
Of course, it would be extraordinary if we could confirm, through
direct detection, the existence of dark matter in our own galaxy. For
that purpose, one should measure its rotation curve, which is much
more difficult because of obscuration by dust in the disk, as well as
problems with the determination of reliable galactocentric distances
for the tracers. Nevertheless, the rotation curve of the Milky Way has
been measured and conforms to the usual picture, with a plateau value
of the rotation velocity of 220 km/s. For dark matter searches, the
crucial quantity is the dark matter density in the solar
neighbourhood, which turns out to be (within a factor of two
uncertainty depending on the halo model) $\rho_{\rm DM} = 0.3$
GeV/cm$^3$. We will come back to direct searched of dark matter in a
later subsection.

\subsection{Baryon fraction in clusters}

Since large clusters of galaxies form through gravitational collapse,
they scoop up mass over a large volume of space, and therefore the ratio
of baryons over the total matter in the cluster should be representative
of the entire universe, at least within a 20\% systematic error. Since
the 1960s, when X-ray telescopes became available, it is known that
galaxy clusters are the most powerful X-ray sources in the
sky.\cite{Sarazin} The emission extends over the whole cluster and
reveals the existence of a hot plasma with temperature $T\sim 10^7 -
10^8$ K, where X-rays are produced by electron bremsstrahlung. Assuming
the gas to be in hydrostatic equilibrium and applying the virial theorem
one can estimate the total mass in the cluster, giving general agreement
(within a factor of 2) with the virial mass estimates. From these
estimates one can calculate the baryon fraction of clusters
\begin{equation}\label{BaryonFraction}
f_{\rm B}h^{3/2} = 0.08 \hspace{5mm} \Rightarrow
\hspace{5mm} {\Omega_B\over\Omega_M} \approx 0.14\,,
\hspace{5mm} {\rm for} \hspace{3mm} h=0.70\,.
\end{equation}
Since $\Omega_{\rm lum} \simeq 0.002 - 0.006$, the previous expression
suggests that clusters contain far more baryonic matter in the form of
hot gas than in the form of stars in galaxies. Assuming this fraction
to be representative of the entire universe, and using the Big Bang
nucleosynthesis value of $\Omega_B = 0.04 \pm 0.01$, for $h=0.7$, we
find
\begin{equation}\label{OmegaXray}
\Omega_M = 0.3 \pm 0.1\ ({\rm statistical})\ 
\pm 20\%\ ({\rm systematic})\,.
\end{equation}
This value is consistent with previous determinations of $\Omega_M$.
If some baryons are ejected from the cluster during gravitational
collapse, or some are actually bound in nonluminous objects like planets,
then the actual value of $\Omega_M$ is smaller than this estimate.

\subsection{Weak gravitational lensing}

Since the mid 1980s, deep surveys with powerful telescopes have
observed huge arc-like features in galaxy clusters. The spectroscopic
analysis showed that the cluster and the giant arcs were at very
different redshifts. The usual interpretation is that the arc is the
image of a distant background galaxy which is in the same line of
sight as the cluster so that it appears distorted and magnified by the
gravitational lens effect: the giant arcs are essentially partial
Einstein rings. From a systematic study of the cluster mass
distribution one can reconstruct the shear field responsible for the
gravitational distortion.\cite{Bartelmann} This analysis shows that
there are large amounts of dark matter in the clusters, in rough
agreement with the virial mass estimates, although the lensing masses
tend to be systematically larger. At present, the estimates indicate
$\Omega_M = 0.2 - 0.3$ on scales $\lsim 6\,h^{-1}$ Mpc.

\subsection{Large scale structure formation and the matter power spectrum}

Although the isotropic microwave background indicates that the
universe in the {\em past} was extraordinarily homogeneous, we know
that the universe {\em today} is far from homogeneous: we observe
galaxies, clusters and superclusters on large scales. These structures
are expected to arise from very small primordial inhomogeneities that
grow in time via gravitational instability, and that may have
originated from tiny ripples in the metric, as matter fell into their
troughs. Those ripples must have left some trace as temperature
anisotropies in the microwave background, and indeed such anisotropies
were finally discovered by the COBE satellite in 1992. However, not
all kinds of matter and/or evolution of the universe can give rise to
the structure we observe today. If we define the density contrast
as
\begin{equation}\label{DensityContrast}
\delta(\vec x,a) \equiv {\rho(\vec x,a)-\bar\rho(a)\over\bar\rho(a)}
= \int d^3\vec{k}\ \delta_k(a)\ e^{i\vec{k}\cdot\vec{x}}\,,
\end{equation}
where $\bar\rho(a)=\rho_0\,a^{-3}$ is the average cosmic density, we need a
theory that will grow a density contrast with amplitude $\delta \sim
10^{-5}$ at the last scattering surface ($z=1100$) up to density
contrasts of the order of $\delta \sim 10^2$ for galaxies at redshifts
$z\ll1$, i.e. today. This is a {\em necessary} requirement for any
consistent theory of structure formation.

Furthermore, the anisotropies observed by the COBE satellite correspond
to a small-amplitude scale-invariant primordial power spectrum of
inhomogeneities
\begin{equation}\label{HarrisonZeldovich}
P(k) = \langle|\delta_k|^2\rangle \propto k^n\,, \hspace{5mm} {\rm with}
\hspace{5mm} n=1\,,
\end{equation}
These inhomogeneities are like waves in the space-time metric. When
matter fell in the troughs of those waves, it created density
perturbations that collapsed gravitationally to form galaxies and
clusters of galaxies, with a spectrum that is also scale invariant.
Such a type of spectrum was proposed in the early 1970s by Edward
R. Harrison, and independently by the Russian cosmologist Yakov
B. Zel'dovich,\cite{HZ} to explain the distribution of
galaxies and clusters of galaxies on very large scales in our
observable universe.

\begin{figure}[t]
\begin{center}
\includegraphics[width=10cm,angle=0]{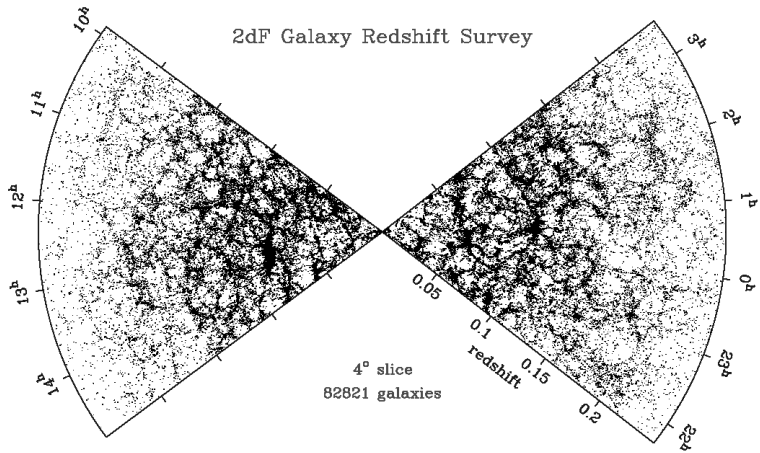}
\end{center}
\caption{The 2 degree Field Galaxy Redshift Survey contains some
250,000 galaxies, cove\-ring a large fraction of the sky up to 
redshifts of $z\leq0.25$.}
\label{2dF}
\end{figure}

Since the primordial spectrum is very approximately represented by a
scale-invariant {\em Gaussian random field}, the best way to present
the results of structure formation is by working with the 2-point
correlation function in Fourier space, the so-called {\em power
spectrum}. If the reprocessed spectrum of inhomogeneities remains
Gaussian, the power spectrum is all we need to describe the galaxy
distribution. Non-Gaussian effects are expected to arise from the
non-linear gravitational collapse of structure, and may be important
at small scales.  The power spectrum measures the degree of
inhomogeneity in the mass distribution on different scales. It depends
upon a few basic ingredientes: a) the primordial spectrum of
inhomogeneities, whether they are Gaussian or non-Gaussian, whether
{\em adiabatic} (perturbations in the energy density) or {\em
isocurvature} (perturbations in the entropy density), whether the
primordial spectrum has {\em tilt} (deviations from scale-invariance),
etc.; b) the recent creation of inhomogeneities, whether {\em cosmic
strings} or some other topological defect from an early phase
transition are responsible for the formation of structure today; and
c) the cosmic evolution of the inhomogeneity, whether the universe has
been dominated by cold or hot dark matter or by a cosmological
constant since the beginning of structure formation, and also
depending on the rate of expansion of the universe.

\begin{figure}[t]
\vspace*{-5mm}
\begin{center}
\includegraphics[width=7.9cm]{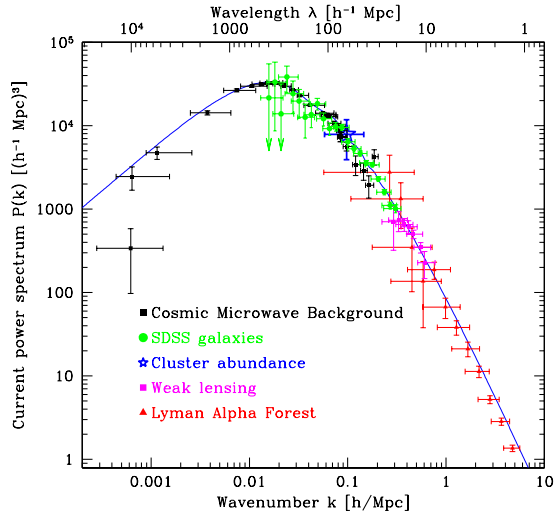}
\end{center}
\vspace*{-0.5cm}
\caption{The measured power spectrum $P(k)$ as a function of
wavenumber $k$. From observations of the Sloan Digital Sky Survey,
CMB anisotropies, cluster abundance, gravitational lensing and
Lyman-$\alpha$ forest.}
\label{fig:Pk}
\end{figure}

The working tools used for the comparison between the observed power
spectrum and the predicted one are very precise N-body numerical
simulations and theoretical models that predict the {\em shape} but not
the {\em amplitude} of the present power spectrum. Even though a large
amount of work has gone into those analyses, we still have large
uncertainties about the nature and amount of matter necessary for
structure formation.  A model that has become a working paradigm is a
flat cold dark matter model with a cosmological constant and
$\Omega_M \sim 0.3$. This model is now been confronted with the recent
very precise measurements from 2dFGRS\cite{2dFGRS} and SDSS.\cite{SDSS}

\subsection{The new redshift catalogs, 2dF and Sloan Digital Sky Survey}

Our view of the large-scale distribution of luminous objects in the
universe has changed dramatically during the last 25 years: from the
simple pre-1975 picture of a distribution of field and cluster
galaxies, to the discovery of the first single superstructures and
voids, to the most recent results showing an almost regular web-like
network of interconnected clusters, filaments and walls, separating
huge nearly empty volumes. The increased efficiency of redshift
surveys, made possible by the development of spectrographs and --
specially in the last decade -- by an enormous increase in
multiplexing gain (i.e. the ability to collect spectra of several
galaxies at once, thanks to fibre-optic spectrographs), has allowed us
not only to do {\em cartography} of the nearby universe, but also to
statistically characterize some of its properties. At the same time,
advances in theoretical modeling of the development of structure, with
large high-resolution gravitational simulations coupled to a deeper
yet limited understanding of how to form galaxies within the dark
matter halos, have provided a more realistic connection of the models
to the observable quantities. Despite the large uncertainties that
still exist, this has transformed the study of cosmology and
large-scale structure into a truly quantitative science, where theory
and observations can progress together.

\begin{figure}[t]
\begin{center}
\includegraphics[width=7.5cm]{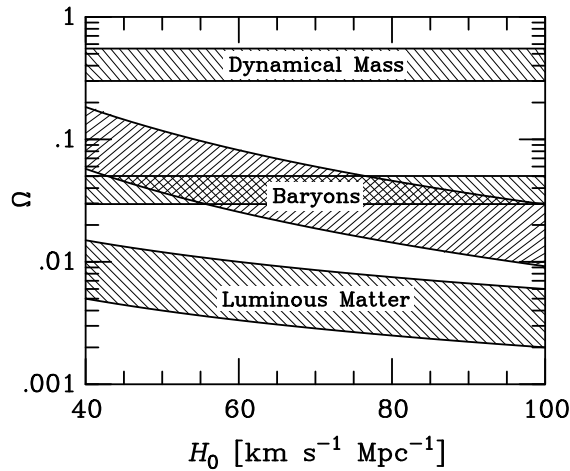} 
\caption{The observed cosmic matter components as functions of the
Hubble expansion parameter. The luminous matter component is given by
$0.002\leq\Omega_{\rm lum}\leq0.006$; the galactic halo component is the
horizontal band, $0.03\leq\Omega_{\rm halo}\leq0.05$, crossing the
baryonic component from BBN, $\Omega_B\,h^2=0.0244\pm0.0024$; and the
dynamical mass component from large scale structure analysis is given by
$\Omega_M=0.3\pm0.1$. Note that in the range $H_0 = 70\pm7$ km/s/Mpc,
there are {\em three} dark matter problems, see the text.}
\label{SDM}
\end{center}
\end{figure}

\subsection{Summary of the matter content}

We can summarize the present situation with Fig.~\ref{SDM}, for
$\Omega_M$ as a function of $H_0$. There are four bands, the luminous
matter $\Omega_{\rm lum}$; the baryon content $\Omega_B$, from BBN;
the galactic halo component $\Omega_{\rm halo}$, and the dynamical
mass from clusters, $\Omega_M$. From this figure it is clear that
there are in fact {\em three} dark matter problems: The first one is
where are 90\% of the baryons? Between the fraction predicted by BBN
and that seen in stars and diffuse gas there is a huge fraction which
is in the form of dark baryons. They could be in small clumps of
hydrogen that have not started thermonuclear reactions and perhaps
constitute the dark matter of spiral galaxies' halos. Note that
although $\Omega_B$ and $\Omega_{\rm halo}$ coincide at $H_0\simeq70$
km/s/Mpc, this could be just a coincidence.  The second problem is
what constitutes 90\% of matter, from BBN baryons to the mass inferred
from cluster dynamics? This is the standard dark matter problem and
could be solved in the future by direct detection of a weakly
interacting massive particle in the laboratory.  And finally, since we
know from observations of the CMB that the universe is flat, the rest,
up to $\Omega_0=1$, must be a diffuse vacuum energy, which affects the
very large scales and late times, and seems to be responsible for the
present acceleration of the universe, see Section 3.

\section{The anisotropies of the microwave background}

One of the most remarkable observations ever made my mankind is the
detection of the relic background of photons from the Big Bang. This
background was predicted by George Gamow and collaborators in the 1940s,
based on the consistency of primordial nucleosynthesis with the observed
helium abundance. They estimated a value of about 10 K, although a
somewhat more detailed analysis by Alpher and Herman in 1950 predicted
$T_\gamma \approx 5$ K. Unfortunately, they had doubts whether the
radiation would have survived until the present, and this remarkable
prediction slipped into obscurity, until Dicke, Peebles, Roll and
Wilkinson studied the problem again in 1965.\cite{Dicke} Before they
could measure the photon background, they learned that Penzias and
Wilson had observed a weak isotropic background signal at a radio
wavelength of 7.35 cm, corresponding to a blackbody temperature of
$T_\gamma=3.5\pm1$ K. They published their two papers back to back, with
that of Dicke et al.  explaining the fundamental significance of their
measurement.

\begin{figure}[t]
\begin{center}
\includegraphics[width=10cm]{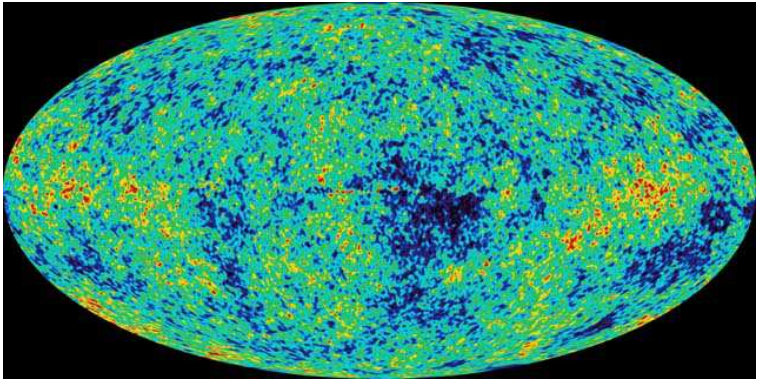} 
\caption{The anisotropies of the microwave background measured by
the WMAP satellite with 10 arcminute resolution. It shows the
intrinsic CMB anisotropies at the level of a few parts in $10^5$.
The galactic foreground has been properly subtracted. The amount of 
information contained in this map is enough to determine most of
the cosmological parameters to few percent accuracy.}
\label{fig:WMAP}
\end{center}
\end{figure}

Since then many different experiments have confirmed the existence of
the microwave background. The most outstanding one has been the Cosmic
Background Explorer (COBE) satellite, whose FIRAS instrument measured
the photon background with great accuracy over a wide range of
frequencies. Nowadays, the photon spectrum is confirmed
to be a blackbody spectrum with a temperature\cite{FIRAS}
\begin{equation}\label{T0}
T_{_{\rm CMB}} = 2.728 \pm 0.002 \ {\rm K} \ 
({\rm systematic}, \ 95\%\ {\rm c.l.}) \ 
\pm 7 \ \mu\!{\rm K} \ (1\sigma\ {\rm statistical})
\end{equation}
In fact, this is the best blackbody spectrum ever measured, with
spectral distortions below the level of 10 parts per million (ppm).
Moreover, the differential microwave radiometer (DMR) instrument on
COBE, with a resolution of about $7^\circ$ in the sky, has also
confirmed that it is an extraordinarily isotropic background.\cite{DMR}
The deviations from isotropy, i.e. differences in the temperature of the
blackbody spectrum measured in different directions in the sky, are of
the order of 20\,$\mu$K on large scales, or one part in $10^5$.  Soon
after COBE, other groups quickly confirmed the detection of temperature
anisotropies at around 30\,$\mu$K and above, at higher multipole numbers
or smaller angular scales. Last year, the satellite Wilkinson Microwave
Anisotropy Probe (WMAP)\cite{WMAP} measured the full sky CMB
anisotropies with 10 arcminutes' resolution, and obtained the most
precise map to date of the microwave sky, see Fig.~\ref{fig:WMAP}

\subsection{Acoustic oscillations in the plasma before recombination}

\begin{figure}[t]
\begin{center}
\includegraphics[width=7cm]{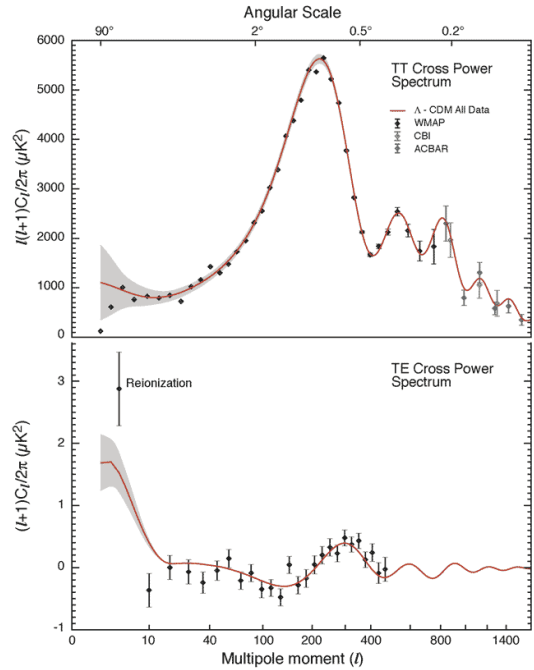} 
\caption{The Angular Power Spectrum of CMB temperature anisotropies,
compared with the cross-correlation of temperature-polarization
anisotropies.}
\label{fig:Power}
\end{center}
\end{figure}

The physics of the CMB anisotropies is relatively simple.\cite{CMB}
The universe just before recombination is a very tightly coupled fluid,
due to the large electromagnetic Thomson cross section $\sigma_T =
8\pi\alpha^2/3m_e^2\simeq0.7$ barn. Photons scatter off charged
particles (protons and electrons), and carry energy, so they feel the
gravitational potential associated with the perturbations imprinted in
the metric during inflation. An overdensity of baryons (protons and
neutrons) does not collapse under the effect of gravity until it enters
the causal Hubble radius. The perturbation continues to grow until
radiation pressure opposes gravity and sets up acoustic oscillations in
the plasma, very similar to sound waves. Since overdensities of the same
size will enter the Hubble radius at the same time, they will oscillate
in phase. Moreover, since photons scatter off these baryons, the
acoustic oscillations occur also in the photon field and induces a
pattern of peaks in the temperature anisotropies in the sky, at
different angular scales, see Fig.~\ref{fig:Power}.

There are three different effects that determine the temperature
anisotropies we observe in the CMB. First, gravity: photons fall in and
escape off gravitational potential wells, characterized by $\Phi$ in the
comoving gauge, and as a consequence their frequency is gravitationally
blue- or red-shifted, $\delta\nu/\nu = \Phi$. If the gravitational
potential is not constant, the photons will escape from a larger or
smaller potential well than they fell in, so their frequency is also
blue- or red-shifted, a phenomenon known as the Rees-Sciama effect.
Second, pressure: photons scatter off baryons which fall into
gravitational potential wells and the two competing forces create
acoustic waves of compression and rarefaction. Finally, velocity:
baryons accelerate as they fall into potential wells. They have minimum
velocity at maximum compression and rarefaction. That is, their velocity
wave is exactly $90^\circ$ off-phase with the acoustic waves. These
waves induce a Doppler effect on the frequency of the photons.
The temperature anisotropy induced by these three effects is therefore
given by\cite{CMB}
\begin{equation}
{\delta T\over T}({\bf r}) = \Phi({\bf r},t_{\rm dec}) + 
2\int_{t_{\rm dec}}^{t_0} \dot\Phi({\bf r},t) dt \ + \
{1\over3}{\delta\rho\over\rho} \ - \ {{\bf r}\cdot{\bf v}\over c}\,.
\label{TemperatureAnisotropy}
\end{equation}
Metric perturbations of different wavelengths enter the horizon at
different times. The largest wavelengths, of size comparable to our
present horizon, are entering now. There are perturbations with
wavelengths comparable to the size of the horizon at the time of last
scattering, of projected size about $1^\circ$ in the sky today, which
entered precisely at decoupling. And there are perturbations with
wavelengths much smaller than the size of the horizon at last
scattering, that entered much earlier than decoupling, all the way to
the time of radiation-matter equality, which have gone through several
acoustic oscillations before last scattering. All these perturbations
of different wavelengths leave their imprint in the CMB anisotropies.

The baryons at the time of decoupling do not feel the gravitational
attraction of perturbations with wavelength greater than the size of the
horizon at last scattering, because of causality.  Perturbations with
exactly that wavelength are undergoing their first contraction, or
acoustic compression, at decoupling. Those perturbations induce a large
peak in the temperature anisotropies power spectrum, see
Fig.~\ref{fig:Power}. Perturbations with wavelengths smaller than these
will have gone, after they entered the Hubble scale, through a series of
acoustic compressions and rarefactions, which can be seen as secondary
peaks in the power spectrum. Since the surface of last scattering is not
a sharp discontinuity, but a region of $\Delta z \sim 100$, there will
be scales for which photons, travelling from one energy concentration to
another, will erase the perturbation on that scale, similarly to what
neutrinos or HDM do for structure on small scales.  That is the reason
why we don't see all the acoustic oscillations with the same amplitude,
but in fact they decay exponentialy towards smaller angular scales, an
effect known as Silk damping, due to photon diffusion.\cite{CMB}

\begin{table}\label{table1}
\caption{{\bf The parameters of the standard cosmological model}. The
standard model of cosmology has about 20 different parameters, needed
to describe the background space-time, the matter content and the
spectrum of metric perturbations. We include here the present range of
the most relevant parameters (with 1$\sigma$ errors), as recently
determined by WMAP, and the error with which the Planck satellite will
be able to determine them in the near future.  The rate of expansion
is written in units of $H=100\,h$ km/s/Mpc}
\begin{tabular}{lllll}
\hline
physical quantity & 
\multicolumn{1}{c}{symbol} &
\multicolumn{1}{c}{WMAP} &
\multicolumn{1}{c}{Planck} \\
\hline
\hline
total density & 
\multicolumn{1}{c}{$\Omega_0$} &
\multicolumn{1}{c}{$1.02\pm0.02$} &
\multicolumn{1}{c}{0.7\%} \\
\hline
baryonic matter & 
\multicolumn{1}{c}{$\Omega_{\rm B}$} &
\multicolumn{1}{c}{$0.044\pm0.004$} &
\multicolumn{1}{c}{0.6\%} \\
\hline
cosmological constant & 
\multicolumn{1}{c}{$\Omega_\Lambda$} &
\multicolumn{1}{c}{$0.73\pm0.04$} &
\multicolumn{1}{c}{0.5\%} \\
\hline
cold dark matter & 
\multicolumn{1}{c}{$\Omega_{\rm M}$} &
\multicolumn{1}{c}{$0.23\pm0.04$} &
\multicolumn{1}{c}{0.6\%} \\
\hline
hot dark matter & 
\multicolumn{1}{c}{$\Omega_\nu h^2$} &
\multicolumn{1}{c}{$<0.0076$ (95\% c.l.)} &
\multicolumn{1}{c}{1\%} \\
\hline
sum of neutrino masses \hspace{0.7cm} & 
\multicolumn{1}{c}{$\sum m_\nu$ (eV)} &
\multicolumn{1}{c}{$<0.23$ (95\% c.l.)} &
\multicolumn{1}{c}{1\%} \\
\hline
CMB temperature & 
\multicolumn{1}{c}{$T_0\ (K)$} &
\multicolumn{1}{c}{$2.725\pm0.002$} &
\multicolumn{1}{c}{0.1\%} \\
\hline
baryon to photon ratio & 
\multicolumn{1}{c}{$\eta\times10^{10}$} &
\multicolumn{1}{c}{$6.1\pm0.3$} &
\multicolumn{1}{c}{0.5\%} \\
\hline
baryon to matter ratio & 
\multicolumn{1}{c}{$\Omega_{\rm B}/\Omega_{\rm M}$} &
\multicolumn{1}{c}{$0.17\pm0.01$} &
\multicolumn{1}{c}{1\%} \\
\hline
spatial curvature & 
\multicolumn{1}{c}{$\Omega_K$} &
\multicolumn{1}{c}{$<0.02$ (95\% c.l.)} &
\multicolumn{1}{c}{0.5\%} \\
\hline
rate of expansion & 
\multicolumn{1}{c}{$h$} &
\multicolumn{1}{c}{$0.71\pm0.03$} &
\multicolumn{1}{c}{0.8\%} \\
\hline
age of the universe & 
\multicolumn{1}{c}{$t_0$ (Gyr)} &
\multicolumn{1}{c}{$13.7\pm0.2$} &
\multicolumn{1}{c}{0.1\%} \\
\hline
age at decoupling & 
\multicolumn{1}{c}{$t_{\rm dec}$ (kyr)} &
\multicolumn{1}{c}{$379\pm8$} &
\multicolumn{1}{c}{0.5\%} \\
\hline
age at reionization & 
\multicolumn{1}{c}{$t_{\rm r}$ (Myr)} &
\multicolumn{1}{c}{$180\pm100$} &
\multicolumn{1}{c}{5\%} \\
\hline
spectral amplitude & 
\multicolumn{1}{c}{$A$} &
\multicolumn{1}{c}{$0.833\pm0.085$} &
\multicolumn{1}{c}{0.1\%} \\
\hline
spectral tilt & 
\multicolumn{1}{c}{$n_{\rm s}$} &
\multicolumn{1}{c}{$0.95\pm0.03$} &
\multicolumn{1}{c}{0.2\%} \\
\hline
spectral tilt variation& 
\multicolumn{1}{c}{$dn_{\rm s}/d\ln k$} &
\multicolumn{1}{c}{$-0.031\pm0.017$} &
\multicolumn{1}{c}{0.5\%} \\
\hline
tensor-scalar ratio & 
\multicolumn{1}{c}{$r$} &
\multicolumn{1}{c}{$<0.71$ (95\% c.l.)} &
\multicolumn{1}{c}{5\%} \\
\hline
reionization optical depth & 
\multicolumn{1}{c}{$\tau$} &
\multicolumn{1}{c}{$0.17\pm0.04$} &
\multicolumn{1}{c}{5\%} \\
\hline
redshift of equality & 
\multicolumn{1}{c}{$z_{\rm eq}$} &
\multicolumn{1}{c}{$3233\pm200$} &
\multicolumn{1}{c}{5\%} \\
\hline
redshift of decoupling & 
\multicolumn{1}{c}{$z_{\rm dec}$} &
\multicolumn{1}{c}{$1089\pm1$} &
\multicolumn{1}{c}{0.1\%} \\
\hline
width of decoupling & 
\multicolumn{1}{c}{$\Delta z_{\rm dec}$} &
\multicolumn{1}{c}{$195\pm2$} &
\multicolumn{1}{c}{1\%} \\
\hline
redshift of reionization & 
\multicolumn{1}{c}{$z_{\rm r}$} &
\multicolumn{1}{c}{$20\pm10$} &
\multicolumn{1}{c}{2\%} \\
\hline
\end{tabular}
\end{table}

From the observations of the CMB anisotropies it is possible to
determine most of the parameters of the Standard Cosmological Model
with few percent accuracy, see Table 1. However, there are many
degeneracies between parameters and it is difficult to disentangle one
from another.  For instance, as mentioned above, the first peak in the
photon distribution corresponds to overdensities that have undergone
half an oscillation, that is, a compression, and appear at a scale
associated with the size of the horizon at last scattering, about
$1^\circ$ projected in the sky today. Since photons scatter off
baryons, they will also feel the acoustic wave and create a peak in
the correlation function. The height of the peak is proportional to
the amount of baryons: the larger the baryon content of the universe,
the higher the peak. The position of the peak in the power spectrum
depends on the geometrical size of the particle horizon at last
scattering. Since photons travel along geodesics, the projected size
of the causal horizon at decoupling depends on whether the universe is
flat, open or closed. In a flat universe the geodesics are straight
lines and, by looking at the angular scale of the first acoustic peak,
we would be measuring the actual size of the horizon at last
scattering. In an open universe, the geodesics are inward-curved
trajectories, and therefore the projected size on the sky appears
smaller. In this case, the first acoustic peak should occur at higher
multipoles or smaller angular scales. On the other hand, for a closed
universe, the first peak occurs at smaller multipoles or larger
angular scales. The dependence of the position of the first acoustic
peak on the spatial curvature can be approximately given by $l_{\rm
peak} \simeq 220\,\Omega_0^{-1/2}$, where $\Omega_0=\Omega_{\rm M} +
\Omega_\Lambda = 1-\Omega_K$. Present observations by WMAP and other
experiments give $\Omega_0=1.00\pm0.02$ at one standard deviation. The
other acoustic peaks occur at harmonics of this, corresponding to
smaller angular scales. Since the amplitude and position of the
primary and secondary peaks are directly determined by the sound speed
(and, hence, the equation of state) and by the geometry and expansion
of the universe, they can be used as a powerful test of the density of
baryons and dark matter, and other cosmological parameters. With the
joined data from WMAP, VSA, CBI and ACBAR, we have rather good
evidence of the existence of the second and third acoustic peaks,
which confirms one of the most important predictions of inflation $-$
the non-causal origin of the primordial spectrum of perturbations $-$,
and rules out cosmological defects as the dominant source of structure
in the universe. Moreover, since the observations of CMB anisotropies
now cover almost three orders of magnitude in the size of
perturbations, we can determine the much better accuracy the value of
the spectral tilt, $n=0.95\pm0.03$, which is compatible with the
approximate scale invariant spectrum needed for structure formation,
and is a prediction of the simplest models of inflation. For the
moment, there seems to be some indication of a running of the spectral
tilt, i.e. a variation of $n$ with scale, $dn(k)/d\ln k=
-0.03\pm0.02$, but it is not significant, most probably induced by the
relatively low quadrupole seen by WMAP.\cite{WMAP}

\begin{figure}[t]
\begin{center}
\includegraphics[width=6cm]{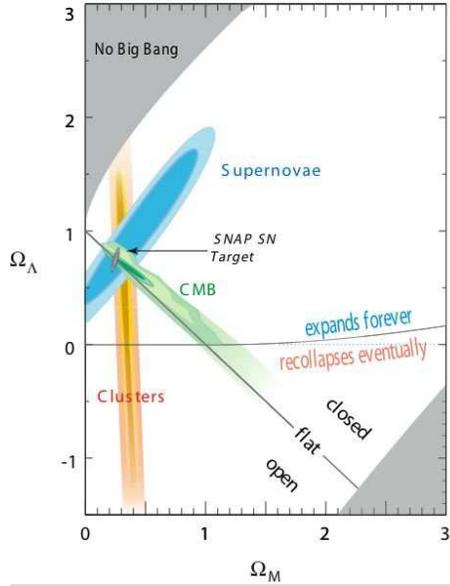} 
\caption{The $(\Omega_M,\,\Omega_\Lambda)$ plane with the present data
set of cosmological observations $-$ the acceleration of the universe,
the large scale structure and the CMB anisotropies $-$ as well as the
future determinations by SNAP and Planck of the fundamental parameters 
which define our Standard Model of Cosmology.}
\label{Future}
\end{center}
\end{figure}

The microwave background has become also a testing ground for theories
of particle physics. In particular, it already gives stringent
constraints on the mass of the neutrino, when analysed together with
large scale structure observations. Assuming a flat $\Lambda$CDM model,
the 2-sigma upper bounds on the sum of the masses of light neutrinos is
$\sum m_\nu < 1.0$ eV for degenerate neutrinos (i.e. without a large
hierachy between them) if we don't impose any priors, and it comes down
to $\sum m_\nu < 0.6$ eV if one imposes the bounds coming from the HST
measurements of the rate of expansion and the supernova data on the
present acceleration of the universe.\cite{Nu} The final bound on the
neutrino density can be expressed as $\Omega_\nu\,h^2 = \sum m_\nu/93.2$
eV $\leq 0.01$.

\section*{Conclusions}

In the last five years we have seen a true revolution in the quality
and quantity of cosmological data that has allowed cosmologists to
determine most of the cosmological parameters with a few percent
accuracy and thus fix a Standard Model of Cosmology. The art of
measuring the cosmos has developed so rapidly and efficiently that one
may be temped of renaming this science as Cosmonomy, leaving the word
Cosmology for the theories of the Early Universe. In summary, we now
know that the stuff we are made of $-$ baryons $-$ constitutes just
about 4\% of all the matter/energy in the Universe, while 25\% is dark
matter $-$ perhaps a new particle species related to theories beyond
the Standard Model of Particle Physics $-$, and the largest fraction,
70\%, some form of diffuse tension also known as dark energy $-$
perhaps a cosmological constant. The rest, about 1\%, could be in the
form of massive neutrinos. 

Nowadays, a host of observations $-$ from CMB anisotropies and large
scale structure to the age and the acceleration of the universe $-$
all converge towards these values, see Fig.~\ref{Future}. Fortunately,
we will have, within this decade, new satellite experiments like
Planck, CMBpol, SNAP as well as deep galaxy catalogs from Earth, to
complement and precisely pin down the values of the Standard Model
cosmological parameters below the percent level, see Table~1.

All these observations would not make any sense without the overall
structure of the inflationary paradigm that determines the homogeneous
and isotropic background on top of which it imprints an approximately
scale invariant gaussian spectrum of adiabatic fluctuations. At
present all observations are consistent with the predictions of
inflation and hopefully in the near future we may have information,
from the polarization anisotropies of the microwave background,
about the scale of inflation, and thus about the physics responsible
for the early universe dynamics.

\section*{Acknowledgments}
It is always a pleasure to attend the Spanish Winter Meetings, but I
want to thank specially the organizers of the 2004 edition in Alicante
for inviting us to an excellent location, with great views and even
better food.  This work was supported in part by a CICYT project
FPA2003-04597.

\end{document}